\documentclass[journal=nalefd,manuscript=letter,manuscript=article]{achemso}
\usepackage[T2A]{fontenc}
\usepackage[utf8]{inputenc}

\usepackage[version=3]{mhchem} 
\usepackage{amsmath,amssymb,bm,upgreek}
\usepackage{placeins} 
\usepackage{alphabeta} 
\usepackage{xspace} 
\usepackage{xcolor}
\usepackage{hyperref}
\usepackage{soul} 

\usepackage{graphicx}
\graphicspath{{./Figures/}}
\usepackage{chemformula}

\usepackage{easyReview}

\renewcommand{\vec}[1]{\bm{#1}}

\author{Gianluca~Gubbiotti}
\affiliation[HZDR]
{CNR-Istituto Officina dei Materiali (IOM), Perugia, Italy}
\email{gubbiotti@iom.cnr.it} 

\author{Olha~Bezsmertna}
\affiliation[HZDR]
{Helmholtz-Zentrum Dresden-Rossendorf e.V., Institute of Ion Beam Physics and Materials Research, 01328 Dresden, Germany}

\author{Oleksandr~Pylypovskyi}
\affiliation[HZDR]
{Helmholtz-Zentrum Dresden-Rossendorf e.V., Institute of Ion Beam Physics and Materials Research, 01328 Dresden, Germany}
\alsoaffiliation[KAU]
{Kyiv Academic University, 03142 Kyiv, Ukraine} 

\author{Rui~Xu}
\affiliation[HZDR]
{Helmholtz-Zentrum Dresden-Rossendorf e.V., Institute of Ion Beam Physics and Materials Research, 01328 Dresden, Germany}

\author{Stéphane~Chiroli}
\affiliation[LSPM]
{LSPM—CNRS, UPR 3407, Université Sorbonne Paris Nord, Villetaneuse, France}
\alsoaffiliation[LAF]
{Laboratoire Albert Fert, UMR 137, CNRS-Thales, Palaiseau, France} 

\author{Fatih~Zighem}
\affiliation[LSPM]
{LSPM—CNRS, UPR 3407, Université Sorbonne Paris Nord, Villetaneuse, France}

\author{Claudia Fernández González}
\affiliation[ALBA]
{Alba Light Source, MISTRAL beamline, Cerdanyola del Vall$\grave{e}$s 08290, Spain}

\author{Andrea~Sorrentino}
\affiliation[ALBA]{Alba Light Source, MISTRAL beamline, Cerdanyola del Vall$\grave{e}$s 08290, Spain}

\author{David~Raftrey}
\affiliation[UC Santa Cruz]
{Department of Physics, University of California, Santa
Cruz, 95064, California, USA}
\alsoaffiliation[BNL]
{Materials Sciences Division, Lawrence Berkeley National Laboratory,
Berkeley, 94720, California, USA}

\author{Daniel~Wolf}
\affiliation[IFW]
{Leibniz Institute for Solid State and Materials Research, 01069 Dresden, Germany}

\author{Axel~Lubk}
\affiliation[IFW]
{Leibniz Institute for Solid State and Materials Research, 01069 Dresden, Germany}

\author{Peter~Fischer}
\affiliation[UC Santa Cruz]
{Department of Physics, University of California, Santa
Cruz, 95064, California, USA}
\alsoaffiliation[BNL]
{Materials Sciences Division, Lawrence Berkeley National Laboratory,
Berkeley, 94720, California, USA}

\author{Damien~Faurie}
\affiliation[LSPM]
{LSPM—CNRS, UPR 3407, Université Sorbonne Paris Nord, Villetaneuse, France}

\author{Denys~Makarov}
\affiliation[HZDR]
{Helmholtz-Zentrum Dresden-Rossendorf e.V., Institute of Ion Beam Physics and Materials Research, 01328 Dresden, Germany}
\email{d.makarov@hzdr.de}

\title[Title]
  {Curvilinear magnonic crystal based on  3D hierarchical nanotemplates}
  

\begin{document}
\begin{abstract}
Curvilinear magnetic nanostructures enable control of magnetization dynamics through geometry-induced anisotropy and chiral interactions as well as magnetic field modulation. In this work, we report a curvilinear magnonic crystal based on large-area square arrays of truncated nanospikes fabricated by conformal coating of 3D hierarchical templates with permalloy thin films. Brillouin light scattering spectroscopy reveals anisotropic band structure with multiple dispersive and folded Bloch-type dispersive spin-wave modes as well as non-dispersive modes exhibiting direction-dependent frequency shifts and intensity asymmetries along lattice principal axes.  Finite element micromagnetic simulations indicate that curvature-induced variations of the demagnetizing field govern the magnonic response, enabling the identification of modes propagating in nanochannels and other localized on nanospike apexes or along the ridges connecting adjacent nanospikes.  The combination of geometric curvature and optical probing asymmetry produces directional dependence of magnonic bands, establishing 3D hierarchical templates as a versatile platform for curvature-engineered magnonics.
\end{abstract}

\maketitle


3D magnonic crystals -- magnetic metamaterials with periodically modulated properties on the nanoscale -- have gained increasing attention due to their potential to revolutionize information processing and wave-based computing~\cite{gubbiotti2019three}. Unlike traditional 1D and 2D magnonic structures~\cite{gubbiotti2010brillouin, tacchi2016brillouin, krawczyk2014review}, 3D architectures offer enhanced control over spin-wave (SW) propagation, richer band structures, novel nonreciprocal and interference phenomena, and engineered SW pathways which are key phenomena to developing energy efficient all-magnon circuits for neuromorphic computing and reconfigurable, multifunctional magnonic devices~\cite{sahoo2021observation, guo2023realization, cheenikundil2025defect, kumar2025magnetic, guo2025coherent, birch2025nanosculpted}. 

Recent activities on 3D magnetic nanostructures were extended to studies of magnonic effects in complex-shaped samples using Brillouin light scattering (BLS) spectroscopy and ferromagnetic resonance (FMR) techniques.
There are numerous theoretical and experimental explorations of SW phenomena in complex architectures like wireframes {\cite{sahoo2018ultrafast, cheenikundil2022high, guo2025coherent}}, nanowires {\cite{gallardo2022high, landeros2022tubular,korber2022curvilinear, brevis2024curvature}}, periodic segments~\cite{sahoo2021observation, dobrovolskiy2022complex, guo2023realization}, waveguides with the broken translational symmetry~\cite{martyshkin2019vertical}, and meander-shaped magnonic crystals~\cite{beginin2018spin, gubbiotti2022spin,sakharov2020spin,gubbiotti2021magnonic}. So far, 3D~magnonics has focused on the design of plane and straight segments as well as networks of interconnected magnonic waveguides, and therefore ignored the effects of non-negligible geometric curvature~\cite{gubbiotti2025roadmap}. 
Curvilinear magnonics explicitly benefits from the effects of geometric curvature like anisotropic and chiral responses in the spirit of curvilinear and 3D magnetism {\cite{dobrovolskiy2022complex,makarov2022new}}. 
Despite being still in an early stage, the topic of curvilinear magnonics has already developed appealing theoretical predictions, yet not accessed experimentally. It has been shown, that finite tangential magnetostatic charges lead to nonreciprocity in SW propagation along a tube~{\cite{Otalora2016curvature,Sheka2020nonlocal,otalora2017assymetric}}. A 1D~curvilinear magnonic crystal designed by a periodically alternating curvature develops band gap edges determined by the effective geometric potentials~\cite{Korniienko2019curvature}. Recent studies reveal the role of the sample topology and a curvature-induced Berry phase in SW dynamics~\cite{dAquino2025nonreciprocal,thonikkadavan2025rotating}. A recent development of magnetic hierarchical nanostructures~\cite{bezsmertna2024magnetic} offers a possibility to extend the experimental framework of the curvilinear magnetism~{\cite{dobrovolskiy2022complex}} on curvilinear magnonics and unlock new functionalities for next-generation spintronic and magnonic devices.

Here, we report the first curvilinear magnonic crystal based on 3D hierarchical nanotemplates consisting of truncated nanospike structures arranged in a square lattice covered by soft magnetic permalloy thin films. We measured the magnonic band structures (i.e. frequency~vs.~wave~vector) by using wave~vector-resolved BLS spectroscopy for two orientations of the externally applied magnetic field along the high symmetry directions, i.e. $[1\,0]$ and $[1\,1]$ directions of the square lattice. The experimental results are supported by numerical simulations, which accurately reproduce the anisotropic magnonic band structure and enable assignment of the spatial profiles and localization characteristics of the experimentally measured SW modes. We identify several intrinsic features of curvilinear magnonic crystals including symmetry-broken excitation of SWs and asymmetric intensity between Stokes and anti-Stokes sides of the spectrum. The excitation range and intensity of the spectrum branches are determined by the curvature profile of geometry. The curvilinear magnonic crystal supports a large amount of localized modes originating from the sharp bends of the geometry. 

\begin{figure}
\includegraphics[width=1\textwidth]{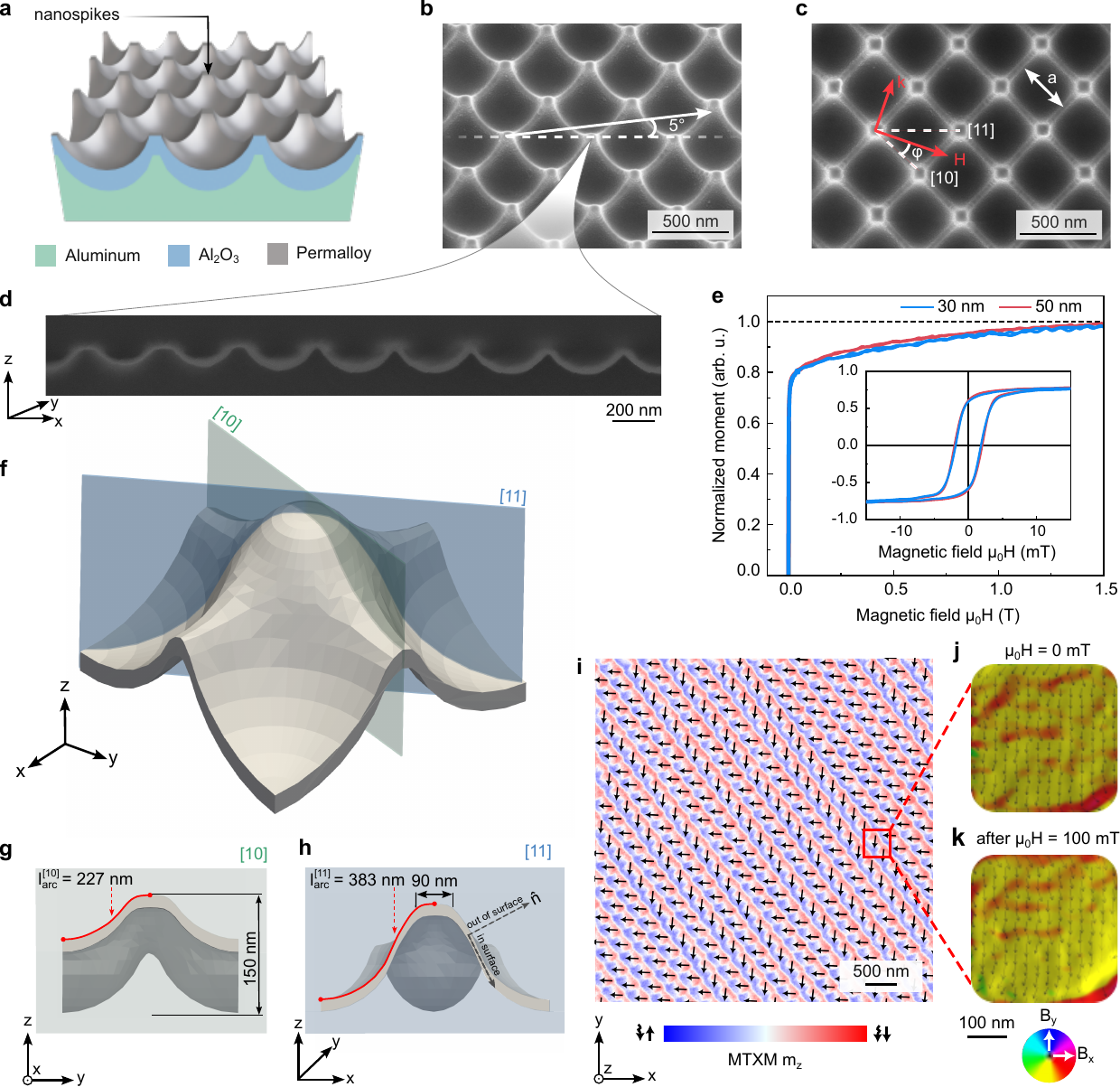}
\caption{(a) Schematics of a curvilinear hierarchical template. Zoomed-in SEM images of (b) tilted and (c) top views of a 30-nm-thick Py film. Schematics in (b) indicates the position of the cross-section shown in (d) that was cut a small~(5$^{\circ}$) tilt angle of milling with respect to the diagonal of the square lattice. (e) $M(H)$ hysteresis loops in an in-plane magnetic field for samples of 30- and 50-nm-thickness. The main panel shows the dependence at high field, low-field hysteresis are shown in the inset. Based on cross-sectional cut (d), the geometry of a single truncated nanospike is reconstructed (f). Parallel (g) and diagonal (h) cross-sections of the reconstructed geometry, showing the lateral dimensions of a single nanospike. (i) A map of the out-of-plane magnetization component measured with MTXM of a 50-nm-thick sample, revealing formation of zigzag-like domains at remanence. (j,k) Mapping of the in-plane components of the projected magnetic induction ($B_x$, $B_y$) within a single nanoindentation between spikes reconstructed by off-axis electron holography: (j) initial remanent state, (k) remanent state after the sample was exposed to a 100~mT out-of-plane magnetic field.}
\label{fig:1}
\end{figure}

We fabricated a curvilinear magnonic crystal using anodized aluminium oxide template method {\cite{xu2022well}}. The aluminium foil is anodized and etched after nanoimprinting that defines a hierarchical square pattern without defects over several cm$^2$ with nanoscale features of 50~nm-wide plateaus and a periodicity of~400~nm. The oxidized substrate is covered by a Permalloy (Ni$_{81}$Fe$_{19}$=Py) layer via magnetron sputtering (Figure~\ref{fig:1}a). Scanning electron microscopy (SEM) images of the final geometry of the sample are shown in Figure~\ref{fig:1}b--d. The obtained curvilinear 30- and 50-nm-thick ferromagnetic membranes form regular square lattices of truncated spikes with a height of 150\,nm separated by hemispherical valleys. The distance between spikes along the $[1\,0]$ ($[1\,1]$) direction is $a = 400$\,nm ($\sqrt{2}a = 566$\,nm). Using focused ion beam (FIB) cut inclined under an angle of $5^\circ$ to the $[1\,1]$ axis of the lattice (Figure~\ref{fig:1}b,d) we reconstruct the geometric structural unit of the lattice (Figure~\ref{fig:1}f--h). The spikes of a lateral size of 90\,nm are connected by smooth 30-nm-wide ridges.The distance between spikes following the curved surface along $[1\,0]$ ($[1\,1]$) direction is  $2l_\text{arc}^{[1\,0]}=454$\,nm ($2l_\text{arc}^{[1\,1]}=746$\,nm) (Figure~\ref{fig:1}g,h). In the following, we refer to the in-plane and out-of-plane directions as the directions within the $xy$-plane and along the $\vec{\hat{z}}$-axis, respectively (Figure~\ref{fig:1}f--h). The in-surface directions are locally tangential to the curvilinear membrane and the out-of-surface direction coincides with the membrane normal $\vec{\hat{n}}$ (Figure~\ref{fig:1}h). 
The {integral magnetic properties of the} curvilinear Py membranes are characterized by vibrating sample magnetomery~(VSM). The in-plane hysteresis loops show a coercivity of about 2.5\,mT~(Figure~\ref{fig:1}e). The two branches of the hysteresis merge above a field of 10~mT, indicating elimination of the domain pattern. The gradual increase in magnetic moment is attributed to a continuous competition of the Zeeman energy and the inhomogeneous shape anisotropy that prevents a strictly uniform magnetization distribution at any field. In particular, the in-surface shape anisotropy \cite{carbou2001mathematical, kohn2005another} results in the magnetic component along $\vec{\hat{z}}$ axis (Figure~\ref{fig:1}f) at the inclined walls of the template. This magnetic state with a gradually varying magnetic moment within the structural unit of the lattice is a specific feature of curvilinear magnonics. Magnetic transmission soft X-ray microscopy (MTXM), which is sensitive to the magnetization component parallel to the the photon propagation direction (thus, to the inclined walls of the membrane), indicates that the remanent state consists of periodic zig-zag stripe domains with $90^\circ$ domain walls between them that follow the geometric features of the lattice (Figure~\ref{fig:1}i). {Off-axis lectron holography at a valley shows that the in-plane magnetization $\vec{M}_{xy}$ keeps a high local uniformity after application of an out-of-plane magnetic field corroborating the dominating impact of shape anisotropy at zero field~(Figure~\ref{fig:1}j,k, Supplementary Figure~3)}. To avoid complications with the analysis of BLS spectra due to the presence of magnetic merons~\cite{bezsmertna2024magnetic}, in the following, we focus on the magnetization dynamics in 30-nm-thick Py membranes.

Due to the absence of conservation of the normal component of the SW vector $k_n\vec{\hat{n}}$ in metallic (opaque) materials{\cite{sandercock1979light}}, a laser beam interacting with a curvilinear magnonic crystal probes SWs with a range of in-surface wave vectors
\begin{equation}\label{eq:k-in-surf}
    k_\text{in-surf}(\vec{r}) = \dfrac{4\pi}{\lambda_{\text{laser}}}  \sin [\theta \pm \theta_0(\vec{r})],
\end{equation}
where $\lambda_\text{laser}$ is the laser wavelength (532\,nm), $\theta$ is the incidence angle with respect to $\vec{\hat{z}}$, $\theta_0(\vec{r}) = \arccos(\vec{\hat{n}}\cdot \vec{\hat{z}})$ is the inclination of the normal $\vec{n}$ at the coordinate $\vec{r}$ (Supplementary Figure~4). However, the wave vector measured in our BLS experiments, $k_\text{nom}$, corresponds to the in-plane projection of $k_\text{in-surf}$, which follows the curvilinear geometry, onto the $xy$~plane. As as result, the projection of the Brillouin zone observed in BLS does not match the intrinsic Brillouin zone experienced by the spin waves travelling along the curvilinear magnonic crystal and is determined by the in-plane periodicity of the lattice. 

The spectra of thermally excited SWs are measured at room temperature using BLS spectroscopy in a backscattering configuration {\cite{carlotti2002magneticproperties}}. A magnetic field $\vec{H}$ is applied in the sample plane perpendicular to the incidence direction of the light that defines the direction of the SW wave vector involved in the scattering process. Consequently, the measurements are conducted in the magnetostatic surface wave configuration, also called the Damon--Eshbach (DE) configuration {\cite{damon1961magnetostatic}}, where the projection $\vec{k}_{xy}$ is oriented perpendicular to the in-plane applied magnetic field~$\vec{H}$. The sample is mounted on a goniometer, enabling rotation around the field direction to vary the incidence angle of light~$\theta$ from 0$^\circ$ to 70$^\circ$. We note that the curvilinear geometry leads to the appearance of shadows for $\theta \gtrsim 30^\circ$ in the detection solid angle for scattering light thus limiting asymmetrically the upper boundary of the range of excited wave vectors according to~Eq.~\eqref{eq:k-in-surf}. In the following, we focus on the analysis of the SW propagation for the external magnetic field applied along the [10] and [11] directions of the square lattice. The experimental data are discussed together with the simulation results. The numerical analysis is based on a finite element method (FEM) that solves the Landau--Lifshitz--Gilbert (LLG) equation in the frequency domain providing eigenfrequencies and spatial mode profiles. 

\begin{figure}
\includegraphics[width=\textwidth]{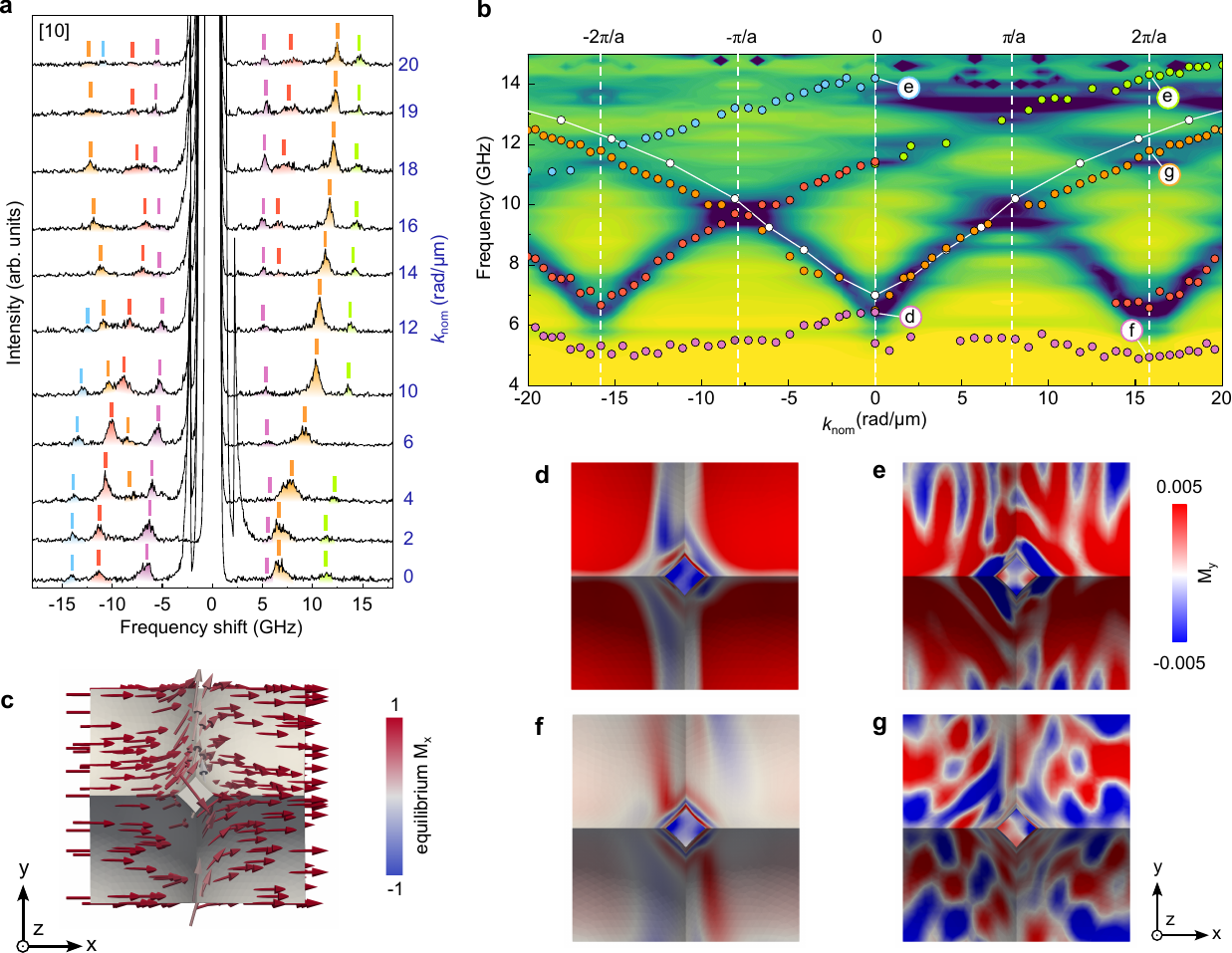}
\caption{(a)~Sequence of measured BLS spectra for different in-plane wave vector $k_\text{nom}$ with magnetic field of~50 mT applied along the $[1\,0]$ direction. Spectra are acquired in the Damon--Eshbach geometry with a 50~mT in-plane magnetic field. Labels near each spectrum correspond to the \mbox{$k_\text{nom}$-value} expressed in rad/$\mu$m. (b)~Experimental SW dispersion measured by BLS (colored symbols) in the same experimental condition of (a). The white symbols correspond to the dispersion of a planar 30-nm-thick reference film. The colored background shows simulation results of the dynamic magnetization. Vertical dashed lines indicate positions of the Brillouin zones. (c)~An equilibrium magnetization distribution over a structural element of the lattice shows a tilt of the magnetization on the inclined facets due to competition between external field and shape anisotropy. The state is simulated in a field of 50\,mT. (d--g)~Spatial distribution of the dynamic magnetization component {$M_{y}$} for the modes labeled (d--g) in panel (b), illustrating the evolution from quasi-uniform to strongly localized oscillations. Low-frequency modes are confined near the apex of truncated nanospikes. Higher-frequency modes extend over the flanks and valleys, with alternating phase between neighboring cells. This is a signature of the formation of a magnonic band through the Bragg reflection.}
\label{fig:2}
\end{figure}

Figure~\ref{fig:2} presents analysis of the SW propagation for a magnetic field $\mu_0 H = 50$\,mT (above coercive field, see Figure~\ref{fig:1}e) applied along the $[1\,0]$ direction. The spectra shown in Figure~\ref{fig:2}a displays well-resolved peaks observed on the Stokes~(S) and anti-Stokes~(AS) sides corresponding to SWs with the opposite $k$-vectors (i.e. positive $k$ for AS and negative $k$ for S sides of the BLS spectra). To follow the evolution of the peak frequencies with wave vector characterizing different SW modes, the peaks have been highlighted by different colors. These modes are displayed with the same color in Figure~\ref{fig:2}b and overlaid with the simulated dynamic magnetization which have been carried out using micromagnetic modelling (see Supplemental Material). The corresponding calculated mode profiles are shown in Figures ~\ref{fig:2}d-g.

As~$k_\text{nom}$~increases, the mode frequencies and the intensities of the peaks evolve progressively yet in a distinct manner on the S~and AS~sides (Figures~\ref{fig:2}a and {~\ref{fig:2}b}). On the S~side, the lowest-frequency resonance (purple peak at around 6\,GHz) shifts monotonically toward lower frequencies as~$k_\text{nom}$ increases up to approximately $12$\,rad/$\upmu$m. For higher~$k_\text{nom}$ values, this mode also appears on the AS~side and becomes nearly dispersionless, i.e. independent of $k_\text{norm}$, with a frequency of about 5\,GHz. Additional peaks (red and light blue) are observed at higher frequencies on the S side of the spectra. These peaks do not have a counterpart on the AS~side and their frequencies decrease with increasing~$k_\text{nom}$. On the AS~side, the most intense peak (orange) can be continuously tracked over the entire~$k_\text{nom}$ range with its frequency increasing as~$k_\text{nom}$ increases.

The mode colored in orange (Figures~\ref{fig:2}a {and ~\ref{fig:2}b}) corresponds to a quasi-uniform precession of the magnetization similar to the DE mode of a planar reference film (white symbols in Figure~\ref{fig:2}b). Although slightly altered by the topography, the quasi-uniform mode extends over the full unit cell of the lattice and remains the most intense and continuous feature in the spectra, appearing predominantly on the S~side due to the surface nonreciprocity inherent to the DE geometry; {its profile is presented in Figure~\ref{fig:2}g}. 

The mode colored in violet (Figures~\ref{fig:2}a,b) is localized at the apex and upper ridges of nanospikes, where the influence of the demagnetizing field is maximum, leading to a weakly dispersive and relatively confined excitation, see the magnetization profile in Figure~\ref{fig:2}c. The simulated mode profile ({see Figure  ~\ref{fig:2}f}) confirms that the precession amplitude is concentrated near the apex of the truncated spike, with nodes appearing progressively along the flanks as the frequency increases. 

The other modes shown in Figures~\ref{fig:2}a,b are altered by the curvilinear profile of the magnetic 3D template. The branch colored in red is reintroduced through band folding and forms a higher-order standing-wave pattern across adjacent nanospikes. It shows a systematic offset between the S~and AS~sides pronounced in the mode intensity. This asymmetry originates from the curvature-induced variation of the optical probing geometry driven by the distribution of $k_\text{in-surf}$ for the given excitation angle. As a result, the S~and AS~signals probe slightly different optical momenta, leading to apparent duplication or lateral shifts of certain branches. The modes colored in green and blue in Figures~\ref{fig:2}a,b have a similar behavior. Their relative intensity and frequency offset vary systematically with the scattering geometry. Because these modes are confined to inclined regions, they are particularly affected by the variation of the local optical wave vector, indicating that curvature couples asymmetries of magnetic and optical responses.

\begin{figure}
\includegraphics[width=\textwidth]{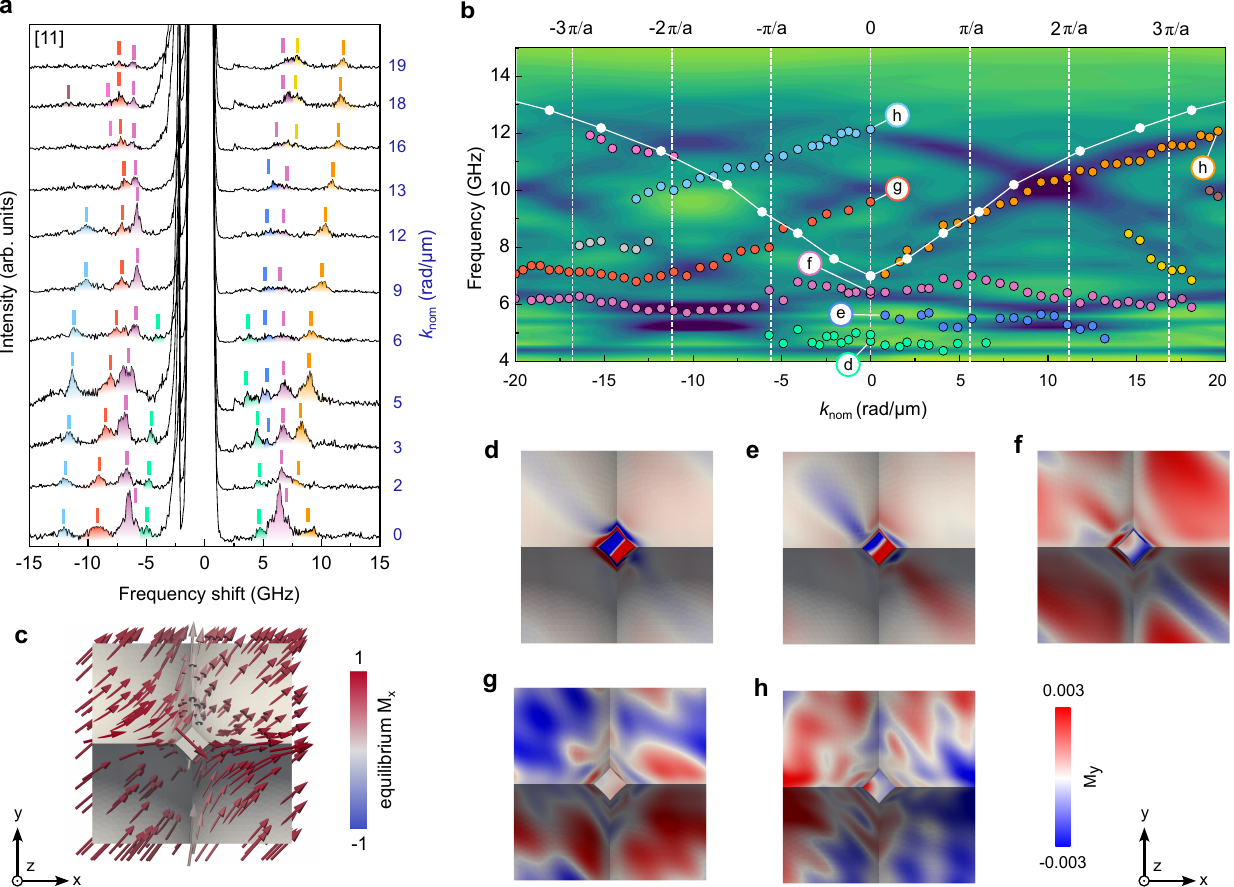}
\caption{(a)~Sequence of measured BLS spectra for different in-plane wave vector $k_\text{nom}$ with magnetic field of 50 mT applied  along the $[1\,1]$ direction. Spectra are acquired in the Damon--Eshbach geometry with a 50~mT in-plane magnetic field. Labels near each spectrum correspond to the~$k_\text{nom}$-value expressed in rad/$\mu$m. (b)~Experimental SW dispersion measured by BLS (colored symbols) along the $[1\,1]$ propagation direction. White symbols correspond to the dispersion of a planar 30-nm-thick reference film. The colored background shows results of micromagnetic simulations, representing the calculated amplitude of the dynamic magnetization $M_{y}$ as a function of frequency and in-plane wave vector $k_\text{nom}$. Vertical dashed lines indicate positions of the Brillouin zones. (c) Simulated equilibrium magnetization configuration $M_{x}$ within one unit cell of the lattice highlighting the gradual reorientation of the magnetization along the inclined facets. (d--h) Spatial maps of the dynamic magnetization component $M_{y}$ for selected modes labeled (d--h) in panel~(b). Low-frequency modes (d,e) are mainly localized at the apex and top edges of the nanospike. Higher-frequency modes (f--h) extend over the lateral flanks and valley regions, forming alternating phase patterns between neighboring cells. This is characteristic of the magnonic band formation and mode hybridization.}
\label{fig:3}
\end{figure}

The analysis of BLS spectra and their comparison with simulations for the case of the SW propagation along $[1\,1]$ direction is given in Figure~\ref{fig:3}. We observe multiple peaks of nonequal intensity confirming that several thermally populated SW modes coexist even close to the center of the Brillouin zone. The S~side exhibits a rich and sharp peak structure with numerous narrow peaks spanning the $6\ldots13$\,GHz range. While the AS~side also displays multiple resonances, those are broader and less intense compared to the peaks on the S side of the spectra. The frequency evolution of the purple, red and blue peaks is similar to what was observed for the $[1\,0]$ direction. Remarkably, their frequencies shift downward by approximately 2\,GHz compared to their values in the $[1\,0]$ orientation. This suggests a strong dependence of the SW dynamics on the magnetic field direction, likely due to modifications in the internal magnetization configuration and the effective field landscape within the sample as well as the curvature-induced coupling with the incoming laser beam. In contrast, there is an additional low-intensity peak shown in green that remains nearly constant at around 5\,GHz.

As follows from Figure~\ref{fig:3}b, the agreement between simulations and experiment is more direct. At $k = 0$, several branches coexist in the $5\ldots 10$\,GHz range, revealing that the lateral surface modulation lifts the degeneracy of the fundamental excitation even at the center of the Brillouin zone. The most intense and dispersive mode indicated in orange (this mode is analogous to the DE mode of the planar film) increases with $k_\text{nom}$ (Figure~\ref{fig:3}b,{h}). As a result of the band folding, this mode continues in the S~side of the spectra (shown in red in Figure~\ref{fig:3}b,g). 

The mode colored in blue (Figure~\ref{fig:3}b,h), which is barely visible experimentally, exhibits alternating phase between neighboring cells, consistent with a weak Bragg folding. Comparing with the $[1\,0]$ case, here the projection of the laser beam on opposite facets is more symmetric leading to a better correspondence between the mode intensities on the S~and AS~sides of the spectra. The observed spectral complexity for the SWs propagating along the $[1\,1]$ direction is thus dominated by the intrinsic magnonic hybridization rather than extrinsic optical geometry effects.


We analyzed the impact of an applied in-plane magnetic field on the mode frequencies ({Supplementary Figure~7}). The magnetic field was applied along the $[1\,0]$ and $[1\,1]$ directions of the square lattice {(Supplementary Figure~7a,b)}. These measurements were performed at a fixed wave vector $|k_\text{nom}| = 4.1$\,rad/$\mu$m, corresponding to an incidence angle of $10^\circ$. For a magnetic field applied along the $[1\,0]$ direction and for positive field values, all the observed peaks in the spectra appear exclusively on one side, regardless of the wave vector $k_\text{nom}$. When the direction of the applied field is reversed, the frequency shifts of all modes also reverse, changing from negative to positive values and \textit{vice versa}. For the propagation along $[1\,0]$ direction, the frequencies of all the observed peaks decrease monotonically starting from $+50$\,mT, reaching a minimum at zero applied field, and subsequently increase as the field is further applied in the opposite direction up to $-50$\,mT {(Supplementary Figure~7c)}. A similar trend is observed when the magnetic field is applied along the $[1\,1]$ direction {(Supplementary Figure~7d)}. In this case, some peaks appear only on one side of the spectra without a corresponding peak on the opposite side. For example, the peak marked by the orange segment is observed only on the AS side of the spectra for positive field and shifts to the S side when the field direction is reversed to negative values.


In summary, we fabricated a curvilinear magnonic crystal of square symmetry based on hierarchical templates with truncated nanospikes connected by narrow ridges. The spin wave dynamics along $[1\,0]$ and $[1\,1]$ directions was analyzed by means of Brillouin light scattering and finite element micromagnetic simulations. A peculiar feature of the geometry is the competition between the external magnetic field and in-surface anisotropy that leads to the periodic modulation of the magnetization distribution even in strong magnetic fields. We identified a pronounced curvature-induced magneto-optical coupling. In addition to common features of 3D magnonic crystals, we found that a continuous variation of the curvature of the sample broadens the range of wave vectors and introduces geometry-driven asymmetry between Stokes and anti-Stokes sides of the spectra. The apex of truncated nanospikes and ridges support dispersiveless modes due to variation of magnetization and related stray fields.  

Our results demonstrate that curvilinear hierarchical templates constitute a scalable platform to tailor magnonic dispersion and control nonreciprocal SW transport. This work opens new opportunities for designing curvature-engineered magnonic devices for prospective low-power and wave-based information processing.

\begin{acknowledgement}
We thank Dr. Ruslan Salikhov (HZDR) for the support with VSM measurements, Dr. Nico Klinger and Andreas Worbs (both HZDR) for their guidance during SEM and FIB operation.
This work was supported in part via the European Union's Horizon Europe Research and Innovation Programme (Grant Agreement No.~101070066; REGO project) and from the ERC grant 3DmultiFerro (Project No. 101141331). D.R. and P.F were supported by the U.S. Department of Energy, Office of Science, Office of Basic Energy Sciences, Materials Sciences and Engineering Division under Contract No. DE-AC02-05-CH11231 (NEMM program MSMAG). A.L. and D.W. acknowledge financial support by the Collaborative Research Center SFB 1143 (project-id 247310070). G.G. acknowledges funding by the European Union–NextGenerationEU, Mission~4, Component~1, under the Italian Ministry of University and Research~(MUR) National Innovation Ecosystem grant ECS00000041-VITALITY-CUP B43C22000470005. These MTXM experiments were performed at the MISTRAL beamline at ALBA Synchrotron with the collaboration of ALBA staff. The ALBA Synchrotron is funded by the Ministry of Research and Innovation of Spain, by the Generalitat de Catalunya and by European FEDER funds. S.C., F.Z., and D.F. acknowledge PEPR SPIN program (France 2030 - SpinTheory project), under ANR grant ANR-22-EXSP-0009. 
\end{acknowledgement}

\clearpage


\begin{mcitethebibliography}{38}
\providecommand*\natexlab[1]{#1}
\providecommand*\mciteSetBstSublistMode[1]{}
\providecommand*\mciteSetBstMaxWidthForm[2]{}
\providecommand*\mciteBstWouldAddEndPuncttrue
  {\def\EndOfBibitem{\unskip.}}
\providecommand*\mciteBstWouldAddEndPunctfalse
  {\let\EndOfBibitem\relax}
\providecommand*\mciteSetBstMidEndSepPunct[3]{}
\providecommand*\mciteSetBstSublistLabelBeginEnd[3]{}
\providecommand*\EndOfBibitem{}
\mciteSetBstSublistMode{f}
\mciteSetBstMaxWidthForm{subitem}{(\alph{mcitesubitemcount})}
\mciteSetBstSublistLabelBeginEnd
  {\mcitemaxwidthsubitemform\space}
  {\relax}
  {\relax}

\bibitem[Gubbiotti(2019)]{gubbiotti2019three}
Gubbiotti,~G., Ed. \emph{Three-dimensional magnonics: layered, micro-and
  nanostructures}; Jenny Stanford Publishing, 2019\relax
\mciteBstWouldAddEndPuncttrue
\mciteSetBstMidEndSepPunct{\mcitedefaultmidpunct}
{\mcitedefaultendpunct}{\mcitedefaultseppunct}\relax
\EndOfBibitem
\bibitem[Gubbiotti \latin{et~al.}(2010)Gubbiotti, Tacchi, Madami, Carlotti,
  Adeyeye, and Kostylev]{gubbiotti2010brillouin}
Gubbiotti,~G.; Tacchi,~S.; Madami,~M.; Carlotti,~G.; Adeyeye,~A.~O.;
  Kostylev,~M. Brillouin light scattering studies of planar metallic magnonic
  crystals. \emph{Journal of Physics D: Applied Physics} \textbf{2010},
  \emph{43}, 264003\relax
\mciteBstWouldAddEndPuncttrue
\mciteSetBstMidEndSepPunct{\mcitedefaultmidpunct}
{\mcitedefaultendpunct}{\mcitedefaultseppunct}\relax
\EndOfBibitem
\bibitem[Tacchi \latin{et~al.}(2016)Tacchi, Gubbiotti, Madami, and
  Carlotti]{tacchi2016brillouin}
Tacchi,~S.; Gubbiotti,~G.; Madami,~M.; Carlotti,~G. Brillouin light scattering
  studies of 2D magnonic crystals. \emph{Journal of Physics: Condensed Matter}
  \textbf{2016}, \emph{29}, 073001\relax
\mciteBstWouldAddEndPuncttrue
\mciteSetBstMidEndSepPunct{\mcitedefaultmidpunct}
{\mcitedefaultendpunct}{\mcitedefaultseppunct}\relax
\EndOfBibitem
\bibitem[Krawczyk and Grundler(2014)Krawczyk, and Grundler]{krawczyk2014review}
Krawczyk,~M.; Grundler,~D. Review and prospects of magnonic crystals and
  devices with reprogrammable band structure. \emph{Journal of Physics:
  Condensed Matter} \textbf{2014}, \emph{26}, 123202\relax
\mciteBstWouldAddEndPuncttrue
\mciteSetBstMidEndSepPunct{\mcitedefaultmidpunct}
{\mcitedefaultendpunct}{\mcitedefaultseppunct}\relax
\EndOfBibitem
\bibitem[Sahoo \latin{et~al.}(2021)Sahoo, May, van Den~Berg, Mondal, Ladak, and
  Barman]{sahoo2021observation}
Sahoo,~S.; May,~A.; van Den~Berg,~A.; Mondal,~A.~K.; Ladak,~S.; Barman,~A.
  Observation of Coherent Spin Waves in a Three-Dimensional Artificial Spin Ice
  Structure. \emph{Nano Letters} \textbf{2021}, \emph{21}, 4629--4635\relax
\mciteBstWouldAddEndPuncttrue
\mciteSetBstMidEndSepPunct{\mcitedefaultmidpunct}
{\mcitedefaultendpunct}{\mcitedefaultseppunct}\relax
\EndOfBibitem
\bibitem[Guo \latin{et~al.}(2023)Guo, Deenen, Xu, Hamdi, and
  Grundler]{guo2023realization}
Guo,~H.; Deenen,~A. J.~M.; Xu,~M.; Hamdi,~M.; Grundler,~D. Realization and
  Control of Bulk and Surface Modes in 3D Nanomagnonic Networks by Additive
  Manufacturing of Ferromagnets. \emph{Advanced Materials} \textbf{2023},
  \emph{35}\relax
\mciteBstWouldAddEndPuncttrue
\mciteSetBstMidEndSepPunct{\mcitedefaultmidpunct}
{\mcitedefaultendpunct}{\mcitedefaultseppunct}\relax
\EndOfBibitem
\bibitem[Cheenikundil \latin{et~al.}(2025)Cheenikundil, d’Aquino, and
  Hertel]{cheenikundil2025defect}
Cheenikundil,~R.; d’Aquino,~M.; Hertel,~R. Defect-sensitive high-frequency
  modes in a three-dimensional artificial magnetic crystal. \emph{npj
  Computational Materials} \textbf{2025}, \emph{11}\relax
\mciteBstWouldAddEndPuncttrue
\mciteSetBstMidEndSepPunct{\mcitedefaultmidpunct}
{\mcitedefaultendpunct}{\mcitedefaultseppunct}\relax
\EndOfBibitem
\bibitem[Kumar \latin{et~al.}(2025)Kumar, Mondal, Pal, Mathur, Scott, Berg,
  Adeyeye, Ladak, and Barman]{kumar2025magnetic}
Kumar,~C.; Mondal,~A.~K.; Pal,~S.; Mathur,~S.; Scott,~J.~R.; Berg,~A. v.~D.;
  Adeyeye,~A.~O.; Ladak,~S.; Barman,~A. Magnetic Charge State Controlled
  Spin-Wave Dynamics in Nanoscale Three-Dimensional Artificial Spin Ice.
  \textbf{2025}, \relax
\mciteBstWouldAddEndPunctfalse
\mciteSetBstMidEndSepPunct{\mcitedefaultmidpunct}
{}{\mcitedefaultseppunct}\relax
\EndOfBibitem
\bibitem[Guo \latin{et~al.}(2025)Guo, Lenz, Gołębiewski, Narkowicz, Lindner,
  Krawczyk, and Grundler]{guo2025coherent}
Guo,~H.; Lenz,~K.; Gołębiewski,~M.; Narkowicz,~R.; Lindner,~J.; Krawczyk,~M.;
  Grundler,~D. Coherent Spin Waves in Curved Ferromagnetic Nanocaps of a
  3D-printed Magnonic Crystal. \textbf{2025}, \relax
\mciteBstWouldAddEndPunctfalse
\mciteSetBstMidEndSepPunct{\mcitedefaultmidpunct}
{}{\mcitedefaultseppunct}\relax
\EndOfBibitem
\bibitem[Birch \latin{et~al.}(2025)Birch, Fujishiro, Belopolski, Mogi, Chiew,
  Yu, Nagaosa, Kawamura, and Tokura]{birch2025nanosculpted}
Birch,~M.~T.; Fujishiro,~Y.; Belopolski,~I.; Mogi,~M.; Chiew,~Y.-L.; Yu,~X.;
  Nagaosa,~N.; Kawamura,~M.; Tokura,~Y. Nanosculpted 3D helices of a magnetic
  Weyl semimetal with switchable nonreciprocity. \textbf{2025}, \relax
\mciteBstWouldAddEndPunctfalse
\mciteSetBstMidEndSepPunct{\mcitedefaultmidpunct}
{}{\mcitedefaultseppunct}\relax
\EndOfBibitem
\bibitem[Sahoo \latin{et~al.}(2018)Sahoo, Mondal, Williams, May, Ladak, and
  Barman]{sahoo2018ultrafast}
Sahoo,~S.; Mondal,~S.; Williams,~G.; May,~A.; Ladak,~S.; Barman,~A. Ultrafast
  magnetization dynamics in a nanoscale three-dimensional cobalt tetrapod
  structure. \emph{Nanoscale} \textbf{2018}, \emph{10}, 9981--9986\relax
\mciteBstWouldAddEndPuncttrue
\mciteSetBstMidEndSepPunct{\mcitedefaultmidpunct}
{\mcitedefaultendpunct}{\mcitedefaultseppunct}\relax
\EndOfBibitem
\bibitem[Cheenikundil \latin{et~al.}(2022)Cheenikundil, Bauer, Goharyan,
  d’Aquino, and Hertel]{cheenikundil2022high}
Cheenikundil,~R.; Bauer,~J.; Goharyan,~M.; d’Aquino,~M.; Hertel,~R.
  High-frequency modes in a magnetic buckyball nanoarchitecture. \emph{APL
  Materials} \textbf{2022}, \emph{10}\relax
\mciteBstWouldAddEndPuncttrue
\mciteSetBstMidEndSepPunct{\mcitedefaultmidpunct}
{\mcitedefaultendpunct}{\mcitedefaultseppunct}\relax
\EndOfBibitem
\bibitem[Gallardo \latin{et~al.}(2022)Gallardo, Alvarado-Seguel, and
  Landeros]{gallardo2022high}
Gallardo,~R.~A.; Alvarado-Seguel,~P.; Landeros,~P. High spin-wave asymmetry and
  emergence of radial standing modes in thick ferromagnetic nanotubes.
  \emph{Physical Review B} \textbf{2022}, \emph{105}, 104435\relax
\mciteBstWouldAddEndPuncttrue
\mciteSetBstMidEndSepPunct{\mcitedefaultmidpunct}
{\mcitedefaultendpunct}{\mcitedefaultseppunct}\relax
\EndOfBibitem
\bibitem[Landeros \latin{et~al.}(2022)Landeros, Otálora, Streubel, and
  Kákay]{landeros2022tubular}
Landeros,~P.; Otálora,~J.~A.; Streubel,~R.; Kákay,~A. \emph{Curvilinear
  Micromagnetism}; Springer International Publishing, 2022; pp 163--213\relax
\mciteBstWouldAddEndPuncttrue
\mciteSetBstMidEndSepPunct{\mcitedefaultmidpunct}
{\mcitedefaultendpunct}{\mcitedefaultseppunct}\relax
\EndOfBibitem
\bibitem[Körber \latin{et~al.}(2022)Körber, Verba, Otálora, Kravchuk,
  Lindner, Fassbender, and Kákay]{korber2022curvilinear}
Körber,~L.; Verba,~R.; Otálora,~J.~A.; Kravchuk,~V.; Lindner,~J.;
  Fassbender,~J.; Kákay,~A. Curvilinear spin-wave dynamics beyond the
  thin-shell approximation: Magnetic nanotubes as a case study. \emph{Physical
  Review B} \textbf{2022}, \emph{106}, 014405\relax
\mciteBstWouldAddEndPuncttrue
\mciteSetBstMidEndSepPunct{\mcitedefaultmidpunct}
{\mcitedefaultendpunct}{\mcitedefaultseppunct}\relax
\EndOfBibitem
\bibitem[Brevis \latin{et~al.}(2024)Brevis, Landeros, Lindner, Kákay, and
  Körber]{brevis2024curvature}
Brevis,~F.; Landeros,~P.; Lindner,~J.; Kákay,~A.; Körber,~L.
  Curvature-induced parity loss and hybridization of magnons: Exploring the
  connection of flat and tubular magnetic shells. \emph{Physical Review B}
  \textbf{2024}, \emph{110}, 134428\relax
\mciteBstWouldAddEndPuncttrue
\mciteSetBstMidEndSepPunct{\mcitedefaultmidpunct}
{\mcitedefaultendpunct}{\mcitedefaultseppunct}\relax
\EndOfBibitem
\bibitem[Dobrovolskiy \latin{et~al.}(2022)Dobrovolskiy, Pylypovskyi, Skoric,
  Fern{\'a}ndez-Pacheco, Van Den~Berg, Ladak, and
  Huth]{dobrovolskiy2022complex}
Dobrovolskiy,~O.~V.; Pylypovskyi,~O.~V.; Skoric,~L.; Fern{\'a}ndez-Pacheco,~A.;
  Van Den~Berg,~A.; Ladak,~S.; Huth,~M. \emph{Curvilinear Micromagnetism: From
  Fundamentals to Applications}; Springer International Publishing, 2022\relax
\mciteBstWouldAddEndPuncttrue
\mciteSetBstMidEndSepPunct{\mcitedefaultmidpunct}
{\mcitedefaultendpunct}{\mcitedefaultseppunct}\relax
\EndOfBibitem
\bibitem[Martyshkin \latin{et~al.}(2019)Martyshkin, Beginin, Stognij, Nikitov,
  and Sadovnikov]{martyshkin2019vertical}
Martyshkin,~A.~A.; Beginin,~E.~N.; Stognij,~A.~I.; Nikitov,~S.~A.;
  Sadovnikov,~A.~V. Vertical Spin-Wave Transport in Magnonic Waveguides With
  Broken Translation Symmetry. \emph{IEEE Magnetics Letters} \textbf{2019},
  \emph{10}, 1--5\relax
\mciteBstWouldAddEndPuncttrue
\mciteSetBstMidEndSepPunct{\mcitedefaultmidpunct}
{\mcitedefaultendpunct}{\mcitedefaultseppunct}\relax
\EndOfBibitem
\bibitem[Beginin \latin{et~al.}(2018)Beginin, Sadovnikov, Sharaevskaya,
  Stognij, and Nikitov]{beginin2018spin}
Beginin,~E.~N.; Sadovnikov,~A.~V.; Sharaevskaya,~A.~Y.; Stognij,~A.~I.;
  Nikitov,~S.~A. Spin wave steering in three-dimensional magnonic networks.
  \emph{Applied Physics Letters} \textbf{2018}, \emph{112}, 122404\relax
\mciteBstWouldAddEndPuncttrue
\mciteSetBstMidEndSepPunct{\mcitedefaultmidpunct}
{\mcitedefaultendpunct}{\mcitedefaultseppunct}\relax
\EndOfBibitem
\bibitem[Gubbiotti \latin{et~al.}(2022)Gubbiotti, Sadovnikov, Sheshukova,
  Beginin, Nikitov, Talmelli, Adelmann, and Ciubotaru]{gubbiotti2022spin}
Gubbiotti,~G.; Sadovnikov,~A.; Sheshukova,~S.~E.; Beginin,~E.; Nikitov,~S.;
  Talmelli,~G.; Adelmann,~C.; Ciubotaru,~F. Spin-wave nonreciprocity and
  formation of lateral standing spin waves in CoFeB/Ta/NiFe meander-shaped
  films. \emph{Journal of Applied Physics} \textbf{2022}, \emph{132}\relax
\mciteBstWouldAddEndPuncttrue
\mciteSetBstMidEndSepPunct{\mcitedefaultmidpunct}
{\mcitedefaultendpunct}{\mcitedefaultseppunct}\relax
\EndOfBibitem
\bibitem[Sakharov \latin{et~al.}(2020)Sakharov, Beginin, Khivintsev,
  Sadovnikov, Stognij, Filimonov, and Nikitov]{sakharov2020spin}
Sakharov,~V.~K.; Beginin,~E.~N.; Khivintsev,~Y.~V.; Sadovnikov,~A.~V.;
  Stognij,~A.~I.; Filimonov,~Y.~A.; Nikitov,~S.~A. Spin waves in meander shaped
  YIG film: Toward 3D magnonics. \emph{Applied Physics Letters} \textbf{2020},
  \emph{117}\relax
\mciteBstWouldAddEndPuncttrue
\mciteSetBstMidEndSepPunct{\mcitedefaultmidpunct}
{\mcitedefaultendpunct}{\mcitedefaultseppunct}\relax
\EndOfBibitem
\bibitem[Gubbiotti \latin{et~al.}(2021)Gubbiotti, Sadovnikov, Beginin, Nikitov,
  Wan, Gupta, Kundu, Talmelli, Carpenter, Asselberghs, Radu, Adelmann, and
  Ciubotaru]{gubbiotti2021magnonic}
Gubbiotti,~G.; Sadovnikov,~A.; Beginin,~E.; Nikitov,~S.; Wan,~D.; Gupta,~A.;
  Kundu,~S.; Talmelli,~G.; Carpenter,~R.; Asselberghs,~I.; Radu,~I.~P.;
  Adelmann,~C.; Ciubotaru,~F. Magnonic Band Structure in Vertical
  Meander-Shaped Co40 Fe40 B20 Thin Films. \emph{Physical Review Applied}
  \textbf{2021}, \emph{15}, 014061\relax
\mciteBstWouldAddEndPuncttrue
\mciteSetBstMidEndSepPunct{\mcitedefaultmidpunct}
{\mcitedefaultendpunct}{\mcitedefaultseppunct}\relax
\EndOfBibitem
\bibitem[Gubbiotti \latin{et~al.}(2025)Gubbiotti, Barman, Ladak, Bran,
  Grundler, Huth, Plank, Schmidt, van Dijken, Streubel, Dobrovoloskiy,
  Scagnoli, Heyderman, Donnelly, Hellwig, Fallarino, Jungfleisch, Farhan,
  Maccaferri, Vavassori, Fischer, Tomasello, Finocchio, Clérac, Sessoli,
  Makarov, Sheka, Krawczyk, Gallardo, Landeros, d’Aquino, Hertel, Pirro,
  Ciubotaru, Becherer, Gartside, Ono, Bortolotti, and
  Fernández-Pacheco]{gubbiotti2025roadmap}
Gubbiotti,~G. \latin{et~al.}  2025 roadmap on 3D nanomagnetism. \emph{Journal
  of Physics: Condensed Matter} \textbf{2025}, \emph{37}, 143502\relax
\mciteBstWouldAddEndPuncttrue
\mciteSetBstMidEndSepPunct{\mcitedefaultmidpunct}
{\mcitedefaultendpunct}{\mcitedefaultseppunct}\relax
\EndOfBibitem
\bibitem[Makarov \latin{et~al.}(2022)Makarov, Volkov, Kákay, Pylypovskyi,
  Budinská, and Dobrovolskiy]{makarov2022new}
Makarov,~D.; Volkov,~O.~M.; Kákay,~A.; Pylypovskyi,~O.~V.; Budinská,~B.;
  Dobrovolskiy,~O.~V. New Dimension in Magnetism and Superconductivity: 3D and
  Curvilinear Nanoarchitectures. \emph{Advanced Materials} \textbf{2022},
  \emph{34}\relax
\mciteBstWouldAddEndPuncttrue
\mciteSetBstMidEndSepPunct{\mcitedefaultmidpunct}
{\mcitedefaultendpunct}{\mcitedefaultseppunct}\relax
\EndOfBibitem
\bibitem[Otálora \latin{et~al.}(2016)Otálora, Yan, Schultheiss, Hertel, and
  Kákay]{Otalora2016curvature}
Otálora,~J.~A.; Yan,~M.; Schultheiss,~H.; Hertel,~R.; Kákay,~A.
  Curvature-Induced Asymmetric Spin-Wave Dispersion. \emph{Physical Review
  Letters} \textbf{2016}, \emph{117}, 227203\relax
\mciteBstWouldAddEndPuncttrue
\mciteSetBstMidEndSepPunct{\mcitedefaultmidpunct}
{\mcitedefaultendpunct}{\mcitedefaultseppunct}\relax
\EndOfBibitem
\bibitem[Sheka \latin{et~al.}(2020)Sheka, Pylypovskyi, Landeros, Gaididei,
  Kákay, and Makarov]{Sheka2020nonlocal}
Sheka,~D.~D.; Pylypovskyi,~O.~V.; Landeros,~P.; Gaididei,~Y.; Kákay,~A.;
  Makarov,~D. Nonlocal chiral symmetry breaking in curvilinear magnetic shells.
  \emph{Communications Physics} \textbf{2020}, \emph{3}\relax
\mciteBstWouldAddEndPuncttrue
\mciteSetBstMidEndSepPunct{\mcitedefaultmidpunct}
{\mcitedefaultendpunct}{\mcitedefaultseppunct}\relax
\EndOfBibitem
\bibitem[Otálora \latin{et~al.}(2017)Otálora, Yan, Schultheiss, Hertel, and
  Kákay]{otalora2017assymetric}
Otálora,~J.~A.; Yan,~M.; Schultheiss,~H.; Hertel,~R.; Kákay,~A. Asymmetric
  spin-wave dispersion in ferromagnetic nanotubes induced by surface curvature.
  \emph{Physical Review B} \textbf{2017}, \emph{95}, 184415\relax
\mciteBstWouldAddEndPuncttrue
\mciteSetBstMidEndSepPunct{\mcitedefaultmidpunct}
{\mcitedefaultendpunct}{\mcitedefaultseppunct}\relax
\EndOfBibitem
\bibitem[Korniienko \latin{et~al.}(2019)Korniienko, Kravchuk, Pylypovskyi,
  Sheka, van~den Brink, and Gaididei]{Korniienko2019curvature}
Korniienko,~A.; Kravchuk,~V.; Pylypovskyi,~O.; Sheka,~D.; van~den Brink,~J.;
  Gaididei,~Y. Curvature induced magnonic crystal in nanowires. \emph{SciPost
  Physics} \textbf{2019}, \emph{7}\relax
\mciteBstWouldAddEndPuncttrue
\mciteSetBstMidEndSepPunct{\mcitedefaultmidpunct}
{\mcitedefaultendpunct}{\mcitedefaultseppunct}\relax
\EndOfBibitem
\bibitem[d’Aquino and Hertel(2025)d’Aquino, and
  Hertel]{dAquino2025nonreciprocal}
d’Aquino,~M.; Hertel,~R. Nonreciprocal Inertial Spin-Wave Dynamics in Twisted
  Magnetic Nanostrips. \emph{Physical Review Letters} \textbf{2025},
  \emph{135}\relax
\mciteBstWouldAddEndPuncttrue
\mciteSetBstMidEndSepPunct{\mcitedefaultmidpunct}
{\mcitedefaultendpunct}{\mcitedefaultseppunct}\relax
\EndOfBibitem
\bibitem[Thonikkadavan \latin{et~al.}(2025)Thonikkadavan, d'Aquino, and
  Hertel]{thonikkadavan2025rotating}
Thonikkadavan,~A.; d'Aquino,~M.; Hertel,~R. Rotating Spin Wave Modes in
  Nanoscale Möbius Strips. \textbf{2025}, \relax
\mciteBstWouldAddEndPunctfalse
\mciteSetBstMidEndSepPunct{\mcitedefaultmidpunct}
{}{\mcitedefaultseppunct}\relax
\EndOfBibitem
\bibitem[Bezsmertna \latin{et~al.}(2024)Bezsmertna, Xu, Pylypovskyi, Raftrey,
  Sorrentino, Fernandez-Roldan, Soldatov, Wolf, Lubk, Schäfer, Fischer, and
  Makarov]{bezsmertna2024magnetic}
Bezsmertna,~O.; Xu,~R.; Pylypovskyi,~O.; Raftrey,~D.; Sorrentino,~A.;
  Fernandez-Roldan,~J.~A.; Soldatov,~I.; Wolf,~D.; Lubk,~A.; Schäfer,~R.;
  Fischer,~P.; Makarov,~D. Magnetic Solitons in Hierarchical 3D Magnetic
  Nanoarchitectures of Nanoflower Shape. \emph{Nano Letters} \textbf{2024},
  \emph{24}, 15774--15780\relax
\mciteBstWouldAddEndPuncttrue
\mciteSetBstMidEndSepPunct{\mcitedefaultmidpunct}
{\mcitedefaultendpunct}{\mcitedefaultseppunct}\relax
\EndOfBibitem
\bibitem[Xu \latin{et~al.}(2022)Xu, Zeng, and Lei]{xu2022well}
Xu,~R.; Zeng,~Z.; Lei,~Y. Well-defined nanostructuring with designable anodic
  aluminum oxide template. \emph{Nature Communications} \textbf{2022},
  \emph{13}\relax
\mciteBstWouldAddEndPuncttrue
\mciteSetBstMidEndSepPunct{\mcitedefaultmidpunct}
{\mcitedefaultendpunct}{\mcitedefaultseppunct}\relax
\EndOfBibitem
\bibitem[Carbou(2001)]{carbou2001mathematical}
Carbou,~G. Thin Layers in micromagnetism. \emph{Mathematical Models and Methods
  in Applied Sciences} \textbf{2001}, \emph{11}, 1529--1546\relax
\mciteBstWouldAddEndPuncttrue
\mciteSetBstMidEndSepPunct{\mcitedefaultmidpunct}
{\mcitedefaultendpunct}{\mcitedefaultseppunct}\relax
\EndOfBibitem
\bibitem[Kohn and Slastikov(2005)Kohn, and Slastikov]{kohn2005another}
Kohn,~R.~V.; Slastikov,~V.~V. Another Thin-Film Limit of Micromagnetics.
  \emph{Archive for Rational Mechanics and Analysis} \textbf{2005}, \emph{178},
  227--245\relax
\mciteBstWouldAddEndPuncttrue
\mciteSetBstMidEndSepPunct{\mcitedefaultmidpunct}
{\mcitedefaultendpunct}{\mcitedefaultseppunct}\relax
\EndOfBibitem
\bibitem[Sandercock and Wettling(1979)Sandercock, and
  Wettling]{sandercock1979light}
Sandercock,~J.~R.; Wettling,~W. Light scattering from surface and bulk thermal
  magnons in iron and nickel. \emph{Journal of Applied Physics} \textbf{1979},
  \emph{50}, 7784--7789\relax
\mciteBstWouldAddEndPuncttrue
\mciteSetBstMidEndSepPunct{\mcitedefaultmidpunct}
{\mcitedefaultendpunct}{\mcitedefaultseppunct}\relax
\EndOfBibitem
\bibitem[Carlotti and Gubbiotti(2002)Carlotti, and
  Gubbiotti]{carlotti2002magneticproperties}
Carlotti,~G.; Gubbiotti,~G. Magnetic properties of layered nanostructures
  studied by means of Brillouin light scattering and the surface
  magneto-optical Kerr effect. \emph{Journal of Physics: Condensed Matter}
  \textbf{2002}, \emph{14}, 8199--8233\relax
\mciteBstWouldAddEndPuncttrue
\mciteSetBstMidEndSepPunct{\mcitedefaultmidpunct}
{\mcitedefaultendpunct}{\mcitedefaultseppunct}\relax
\EndOfBibitem
\bibitem[Damon and Eshbach(1961)Damon, and Eshbach]{damon1961magnetostatic}
Damon,~R.; Eshbach,~J. Magnetostatic modes of a ferromagnet slab. \emph{Journal
  of Physics and Chemistry of Solids} \textbf{1961}, \emph{19}, 308--320\relax
\mciteBstWouldAddEndPuncttrue
\mciteSetBstMidEndSepPunct{\mcitedefaultmidpunct}
{\mcitedefaultendpunct}{\mcitedefaultseppunct}\relax
\EndOfBibitem
\end{mcitethebibliography}

\providecommand{\latin}[1]{#1}
\makeatletter
\providecommand{\doi}
  {\begingroup\let\do\@makeother\dospecials
  \catcode`\{=1 \catcode`\}=2 \doi@aux}
\providecommand{\doi@aux}[1]{\endgroup\texttt{#1}}
\makeatother
\providecommand*\mcitethebibliography{\thebibliography}
\csname @ifundefined\endcsname{endmcitethebibliography}
  {\let\endmcitethebibliography\endthebibliography}{}

\end{document}